\documentclass[onecolumn]{elsarticle}%
\usepackage{amssymb}
\usepackage{amsmath}
\usepackage{amsfonts}
\usepackage{graphicx}%
\journal{Journal of Solid State Chemistry}
\begin{document}
\begin{frontmatter}
%% Title, authors and addresses
%% use the tnoteref command within \title for footnotes;
%% use the tnotetext command for theassociated footnote;
%% use the fnref command within \author or \address for footnotes;
%% use the fntext command for theassociated footnote;
%% use the corref command within \author for corresponding author footnotes;
%% use the cortext command for theassociated footnote;
%% use the ead command for the email address,
%% and the form \ead[url] for the home page:
%% \title{Title\tnoteref{label1}}
%% \tnotetext[label1]{}
%% \author{Name\corref{cor1}\fnref{label2}}
%% \ead{email address}
%% \ead[url]{home page}
%% \fntext[label2]{}
%% \cortext[cor1]{}
%% \address{Address\fnref{label3}}
%% \fntext[label3]{}
\title{Manganite charge and orbitally ordered and disordered states probed by Fe substitution into Mn site in LnBaMn$_{1.96}$Fe$%
_{0.04}$O$_{5}$ , LnBaMn$_{1.96}$Fe$_{0.04}$O$_{6}$ and LnBaMn$_{1.96}$Fe$_{0.04}$O$_{5.5}$(Ln=Y, Gd, Sm, Nd, Pr, La)}
%% use optional labels to link authors explicitly to addresses:
%% \author[label1,label2]{}
%% \address[label1]{}
%% \address[label2]{}
\author[a,b]{Alexandre I. Rykov}
\ead{rykov@woody.ocn.ne.jp}
\author[b]{Yutaka Ueda}
\author[a]{ and Kiyoshi Nomura}
\address[a]{School of Engineering, The University of Tokyo, Hongo 7-3-1, 113-8656 Japan}
\address[b]{Institute for Solid State Physics, University of Tokyo, 5-1-5 Kashiwanoha, Kashiwa, Chiba 277-8581, Japan}
\begin{abstract}
The layered manganese oxides LnBaMn$_{1.96}$Fe$_{0.04}$O$_{y}$ (Ln=Y, Gd, Sm, Nd, Pr, La) have been synthesized for $y=5, 5.5$ and $6$. In the oxygen-saturated state $(y=6)$ they exhibit the charge and orbital order at ambient temperature for Ln=Y, Gd, Sm, but unordered $e_{g}$-electronic system for Ln=La,Pr,Nd. Fourfold increase of quadrupole splitting was observed owing to the charge and orbital ordering. This is in agreement with the jumplike increase in distortion of the reduced perovskite-like cell for the charge and orbitally ordered manganites compared to the unordered ones. Substitution of 2 percents of Mn by Fe suppresses the temperatures of structural and magnetic transitions by $20$ to $50$ K. Parameters of the crystal lattices and the room-temperature M\"{o}ssbauer spectra were studied on forty samples whose structures were refined within five symmetry groups: P4/mmm, P4/nmm, Pm-3m, Icma and P2/m. Overwhelming majority of the Fe species are undifferentiated in the M\"{o}ssbauer spectra for most of the samples. Such the single-component spectra in the two-site structures are explained by the preference of Fe towards the site of Mn(III) and by the segmentation of the charge and orbitally ordered domains.
\end{abstract}
\begin{keyword}
manganite \sep charge order \sep layered perovskites \sep orbital order \sep quadrupole interactions
%% PACS codes here, in the form: \PACS code \sep code
%% MSC codes here, in the form: \MSC code \sep code
%% or \MSC[2008] code \sep code (2000 is the default)
\end{keyword}
\end{frontmatter}

\section{Introduction}

The manganites LnBaMn$_{2}$O$_{y}$ ($y=5$ and $6$, Ln=Y and rare-earth
elements) compose a novel family of oxides with the layered structures derived
of perovskite via unmixing the Ln and Ba cations from the A-site into separate
layers\cite{MCDRS,Millange,NKU}. The layered structures for the oxygen
contents $y=6$ and $y=5$ differ by one layer, which is, respectively, LnO and
Ln$\square$, with $\square$ standing for vacant oxygen site. Both these
structures enclose the ions of manganese in the states of half-doping mixed
valence. The Jahn-Teller (JT) ions Mn$^{3+}$ are involved into the planar
checkerboard arrangements either with Mn$^{2+}$ or with Mn$^{4+}$ ions for
$y=5$ and $6$, respectively. The layered arrangement of Ln (III) and Ba(II)
generates on Mn$^{3+}$ cations the staggered systems of either the
out-of-plane d$_{z^{2}}$ orbitals ($y=5$) or the in-plane $d_{3x^{2}-r^{2}%
}/d_{3y^{2}-r^{2}}$ orbitals ($y=6$). The charge ordering between Mn$^{3+}$
and Mn$^{2+}$(Mn$^{4+}$) and orbital ordering between d$_{z^{2}}$
($d_{3x^{2}-r^{2}}$ and $d_{3y^{2}-r^{2}}$) are typically coupled into a bound
process giving birth to the charge-orbitally ordered state denoted hereof as
COO. There occur within a family of the ordered manganites several ways of
stacking these e$_{g}$-electronic sheets into the 3D COO
structures\cite{NKIOYU,Aka}.

On the other hand, the oxygen content $y=5.5$ yields the single-valence
Mn(III) system with no charge differentiation, but with the order of the
Mn$^{3+}$ orbits related to the ordered system of the oxygen
vacancies\cite{Caig,Perca}. The manganite LnBaMn$_{2}$O$_{5.5}$ encloses the
JT Mn$^{3+}$ ions in the equipopulated pyramidal and octahedral coordinations.
The pyramids and octahedra form the chains, in which the orbital order
consists of the alternating parallel and perpendicular to the chains
$d_{x^{2}}$ and $d_{z^{2}}$orbitals filled in octahedra and in pyramids, respectively.

In this work, we employ the M\"{o}ssbauer spectroscopy to explore the variety
of the structural and chemical states adopted by the Fe impurity doped into
the Mn sites for the ordered structures corresponding to $y=5$, $5.5$ and $6$.
For $y=$ $6$, in addition to the A-site ordered layered structures, the cubic
disordered Fe doped systems were also synthesized. We show that the phase
transition succession in LnBaMn$_{1.96}$Fe$_{0.04}$O$_{6} $ remains quite
resembling to that in undoped LnBaMn$_{2}$O$_{6}$.

M\"{o}ssbauer spectroscopy senses the local surrounding of the probe nuclei
($^{57}$Fe in our case) substituted into Mn sites through observing the
electric field gradient (EFG) and magnetic hyperfine fields generated by
adjacent electrons. Based on interpreting these quantities our study reveals
the way by which the impurity ion controls the composition of its surrounding.
In the charge-ordered state, only one component Fe(III) is displayed in
M\"{o}ssbauer spectra. The apparent controversy between the singular Fe$^{3+}$
states and the persistent COO of Mn$^{2+}$/Mn$^{3+}$ in LnBaMn$_{1.96}%
$Fe$_{0.04}$O$_{5}$ and of Mn$^{3+}$/Mn$^{4+}$ in LnBaMn$_{1.96}$Fe$_{0.04}%
$O$_{6}$ can be resolved supposing that the Fe dopants break the long range
COO and shorten the COO correlation lengths.

Three groups of the oxygen-saturated manganites were specified previously
according to the size of Ln \cite{JPSJ2002} as Ln(1$^{\text{st}}$) $=$(La, Pr,
Nd), Ln(2$^{\text{nd}}$)=(Sm, Eu, Gd) and Ln(3$^{\text{rd}}$)=(Y, Tb, Dy, Ho).
A few members from each of these families were investigated in this work for
the effect of the Fe species on the transitions manifested in magnetic
properties. It was shown in several previous works\cite{NYU,Tr} that besides
the A-site disordered and fully ordered systems the partly A-site ordered
layered manganites can also be prepared in special conditions. In this work,
we avoided making the partially A-site ordered samples. Our samples were
stoichiometric and having an integer or half-integer oxygen index exept a few
samples among the A-site disordered systems ([Ar-2] samples), as discussed below.

\section{Experimental}

The manganites LnBaMn$_{1.96}$Fe$_{0.04}$O$_{y}$ were prepared for Ln=Y, Gd,
Sm, (Sm$_{0.9}$Nd$_{0.1}$), (Sm$_{0.1}$Nd$_{0.9}$), Nd, Pr and La. The oxides
Fe$_{2}$O$_{3}$ and Ln$_{2}$O$_{3}$ (Pr$_{6}$O$_{11}$ for Pr-based manganites)
were mixed with the carbonates BaCO$_{3}$ and MnCO$_{3}$. The mixtures were
first annealed in 6N pure Ar (99.9999\%) flow at 1350$^{o}$C. Rapid cooling
from this temperature in Ar has led to obtaining the oxygen-depleted phases
LnBaMn$_{1.96}$Fe$_{0.04}$O$_{5}$. A fraction of each sample was picked out
and denoted as [Ar-1]. Remaining part of a sample was subjected to further
thermal treatments.

\bigskip

Table 1. Symmetry modifications of LnBaMn$_{1.96}$Fe$_{0.04}$O$_{y}$ (Ln=Y and
rare earths from La to Gd) best suited to our x-ray diffraction data according
to their Rietveld analysis. Sequence of annealing steps applied to synthesize
these modifications consisted of a number of gas-heat treatments. Their
conditions are shown in the upper rows. The annealing steps were applied
sequentially. After each step a fraction of sample was picked out. The
fractions labeled [Ar-1], [O-1], [Dis-O], [Ar-2] and [O-2] were obtained after
the steps 1,2,3,4, and 5, respectively. Several samples among the [Ar-2] and
[O-2] series showed a presence of second phase. In these cases, the percentage
of phases provided by FULLPROF program is given.%

\begin{tabular}
[c]{|l|l|l|l|l|l|}\hline
Anneal &  &  &  &  & \\
step N$%
%TCIMACRO{\U{ba}}%
%BeginExpansion
{{}^o}%
%EndExpansion
$ & 1 & 2 & 3 & 4 & 5\\\hline
Gas & Ar(5N) & O$_{2}$ & 95\%O$_{2}$ & Ar(5N) & O$_{2}$\\\hline
T & 1350$^{o}$C & 500$^{o}$C & 1400$^{o}$C & 1450$^{o}$C & 350$^{o}$C\\\hline
Time & 48 h & 48 h & 24 h & 24 h & 100 h\\\hline\hline
Steps & \ \ \ 1 & 1, 2 & 1, 2, 3 & 1, 2, 3, 4 & \ 1, 2, 3, 4, 5\\
Sample & [Ar-1] & [O-1] & [Dis-O] & \ \ \ \ \ [Ar-2] &
\ \ \ \ \ \ [O-2]\\\hline\hline
La & P4/nmm & P4/mmm & P m $\overline{3}$ m & P m $\overline{3}$ m & P m
$\overline{3}$ m\\\hline
Pr & P4/nmm & P4/mmm & P m $\overline{3}$ m & P m $\overline{3}$ m & P m
$\overline{3}$ m\\\hline
Nd & P4/nmm & P4/mmm & P m $\overline{3}$ m & P m $\overline{3}$ m & P m
$\overline{3}$ m\\\hline
Sm$_{0.1}$ & P4/nmm & P4/mmm & P m $\overline{3}$ m & P m $\overline{3}$ m & P
m $\overline{3}$ m\\
Nd$_{0.9}$ &  &  &  &  & \\\hline\hline
Sm$_{0.9}$ & P4/nmm & P4/mmm & P m $\overline{3}$ m & Icma(96\%)+ &
P4/mmm(56\%)\\
Nd$_{0.1}$ &  &  &  & P4/nmm(4\%) & P4/mmm(44\%)\\\hline
Sm & P4/nmm & P4/mmm & P m $\overline{3}$ m & Icma(93\%)+ & P4/mmm(51\%)\\
&  &  &  & P4/nmm(7\%) & P4/mmm(49\%)\\\hline
Gd & P4/nmm & P4/mmm & P m $\overline{3}$ m & Icma(89\%)+ & P4/mmm\\
&  &  &  & P4/nmm(11\%) & \\\hline\hline
Y & P4/nmm & P$\overline{1}$ & Binary & Icma & P$\overline{1}$\\
&  &  & oxides &  & \\\hline
\end{tabular}

\texttt{\smallskip}\medskip

Five protocols of thermal treatment were applied. Sequentially, after each
annealing step a fraction of the sample was picked out, and the remaining part
of the sample was subjected to the next step. Five series of the samples were
obtained in this way. In Table 1, they correspond to columns labeled as
[Ar-1], [O-1], [Dis-O], [Ar-2] and [O-2] according to the final annealing
step. Eight rows of Table 1 correspond to a particular rare-earth Ln. In
total, forty (5$\times$8) samples were analyzed. All the samples of series
[Ar-1], [O-1], and most of the samples from the [Ar-2] and [O-2] series showed
single-phase x-ray patterns (Fig.1). Only a few samples among the [Ar-2] and
[O-2] series showed the presence of a second phase. Members of the [Dis-O]
series were all cubic single-phase, except the Y-based sample. The latter
consisted of binary hexagonal oxides YMnO$_{3}$ (PDF 25-1079) and BaMnO$_{3}%
$(PDF 14-274), and contained no perovskite phase.

Lattice parameters were refined through the analysis of full-profile x-ray
diffraction intensities using FULLPROF program\cite{DBW,Full}. The data were
obtained by means of a "Mac Science" diffractometer using the Cu-$K_{\alpha} $
radiation ($\lambda$= 0.15405 nm\ and 0.15443 nm). Parameters of preferred
orientation of platy crystallites along the axis [001] were refined using
March-Dollase function \cite{Dollase}.

Measurements of magnetization were performed using a SQUID magnetometer in an
applied field of 1 kOe at heating the samples from 5 K to $T_{\text{max}}$ and
then at cooling from $T=T_{\text{max}}$ down to 5 K. This measurement protocol
was applied in LnBaMn$_{1.96}$Fe$_{0.04}$O$_{6}$ for Ln=Sm with $T_{\text{max}%
}$=400 K and for Ln=(Nd$_{0.9}$Sm$_{0.1}$) with $T_{\text{max}}=$370 K. The
magnetization in YBaMn$_{1.96}$Fe$_{0.04}$O$_{6}$ was measured first at
heating from ambient temperature to $T_{\text{max}}$= 600 K (Ln=Y) and then at
cooling from $T=T_{\text{max}}$ down to 5 K. The sample was then remagnetized
at 5K by setting the external field $H=0\pm0.01$ kOe followed by reapplying
$H=1$ kOe. Finally, this sample magnetization was measured at heating up to
300 K.

M\"{o}ssbauer spectra were measured at room temperature and at 11 K. Isomer
shifts are referred relatively $\alpha-$Fe.%

%TCIMACRO{\FRAME{ftbpFU}{2.7959in}{8.1612in}{0pt}{\Qcb{X-ray diffraction
%patterns of Sm$_{0.9}$Nd$_{0.1}$BaMn$_{1.96}$Fe$_{0.04}$O$_{y}$, obtained by
%thermal treatments [Ar-1], [O-1], [Dis-O], [Ar-2] and [O-2] explicated in
%Table 1. Dots represent the observed profiles; solid lines represent
%calculated profiles and difference.}}{\Qlb{f1}}{fi1.eps}%
%{\special{ language "Scientific Word";  type "GRAPHIC";
%maintain-aspect-ratio TRUE;  display "USEDEF";  valid_file "F";
%width 2.7959in;  height 8.1612in;  depth 0pt;  original-width 7.4218in;
%original-height 22.0819in;  cropleft "0";  croptop "1";  cropright "1";
%cropbottom "0";  filename 'Fi1.eps';file-properties "XNPEU";}}}%
%BeginExpansion
\begin{figure}
[ptb]
\begin{center}
\includegraphics[
height=8.1612in,
width=2.7959in
]%
{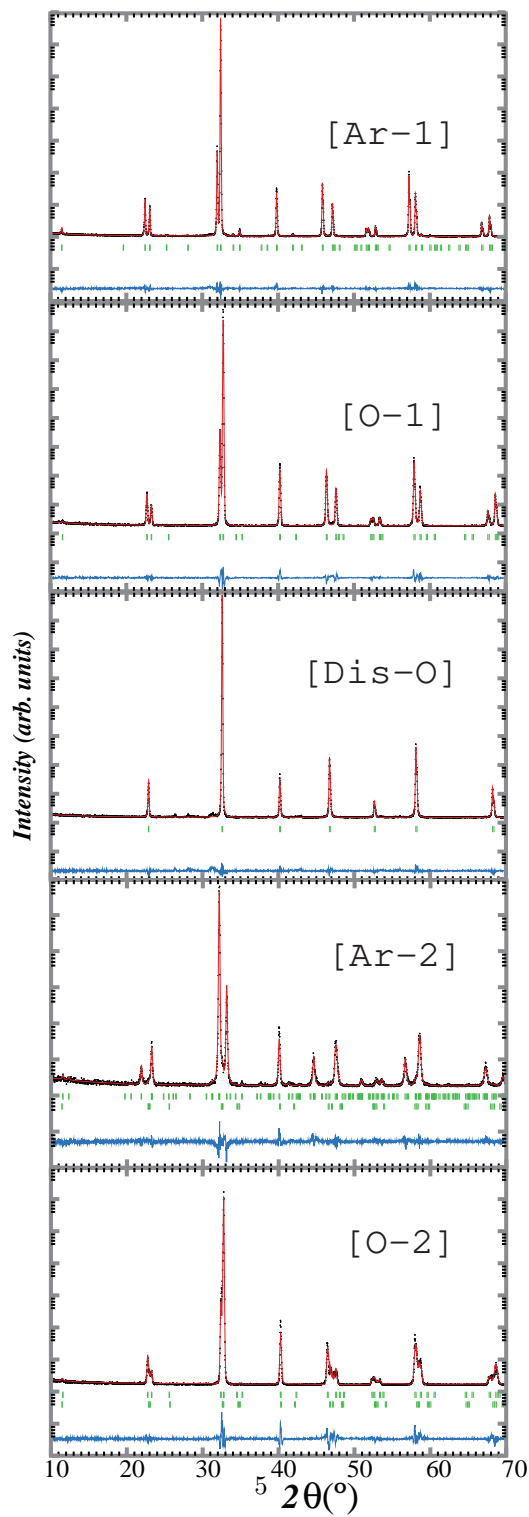}%
\caption{X-ray diffraction patterns of Sm$_{0.9}$Nd$_{0.1}$BaMn$_{1.96}%
$Fe$_{0.04}$O$_{y}$, obtained by thermal treatments [Ar-1], [O-1], [Dis-O],
[Ar-2] and [O-2] explicated in Table 1. Dots represent the observed profiles;
solid lines represent calculated profiles and difference.}%
\label{f1}%
\end{center}
\end{figure}
%EndExpansion

\section{Results and Discussion}

The rest of this article is organized as follows. First, the structure of the
samples obtained is described following the sequence of thermal treatments
[Ar-1], [O-1], [Dis-O], [Ar-2], and [O-2]. It will be shown that the Ln's are
grouped into three groups according to the behaviors of the Ln-based samples
for some of these treatments. Second, these three groups will be characterized
by their M\"{o}ssbauer spectra.

\subsection{Structural Considerations: Phases and Phase Transitions from X-ray
Diffraction and Magnetization}

Three groups of manganites specified previously according to the size of Ln
\cite{JPSJ2002} show a distinct behavior with respect to the thermal
treatments at the steps 3, 4 and 5, as is illustrated by our Table 1. Because
we used for the synthesis of the [Ar-1], [O-1] and [Dis-O] series the
conditions similar to previously reported\cite{Millange,Caig,NKU}, we may
attribute three groups of different behaviors in Table 1 not only to Fe-doped,
but also to clean undoped manganites. Stability of the cubic disordered phase
Pm$\overline{3}$m turns out to depend on the size of Ln.%

%TCIMACRO{\FRAME{ftbpFU}{4.1572in}{5.7648in}{0pt}{\Qcb{The crystal structures
%and symmetry groups employed in Rietveld analysis of x-ray diffraction
%profiles for layer-ordered LnBaMn$_{2}$O$_{5}$(a,b), LnBaMn$_{2}$O$_{6}$(c),
%disordered Ln$_{0.5}$Ba$_{0.5}$MnO$_{3}$ (d), YBaMn$_{2}$O$_{6}$(e), and
%LnBaMn$_{2}$O$_{5.5}$(f). Here Ln = La, Pr, Nd, (Nd$_{0.9}$Sm$_{0.1}$),
%(Nd$_{0.1}$Sm$_{0.9}$), Sm and Gd.}}{\Qlb{f2}}{fi2finalrev.eps}%
%{\special{ language "Scientific Word";  type "GRAPHIC";
%maintain-aspect-ratio TRUE;  display "USEDEF";  valid_file "F";
%width 4.1572in;  height 5.7648in;  depth 0pt;  original-width 6.8554in;
%original-height 9.5441in;  cropleft "0";  croptop "1";  cropright "1";
%cropbottom "0";  filename 'Fi2finalRev.eps';file-properties "XNPEU";}}}%
%BeginExpansion
\begin{figure}
[ptb]
\begin{center}
\includegraphics[
height=5.7648in,
width=4.1572in
]%
{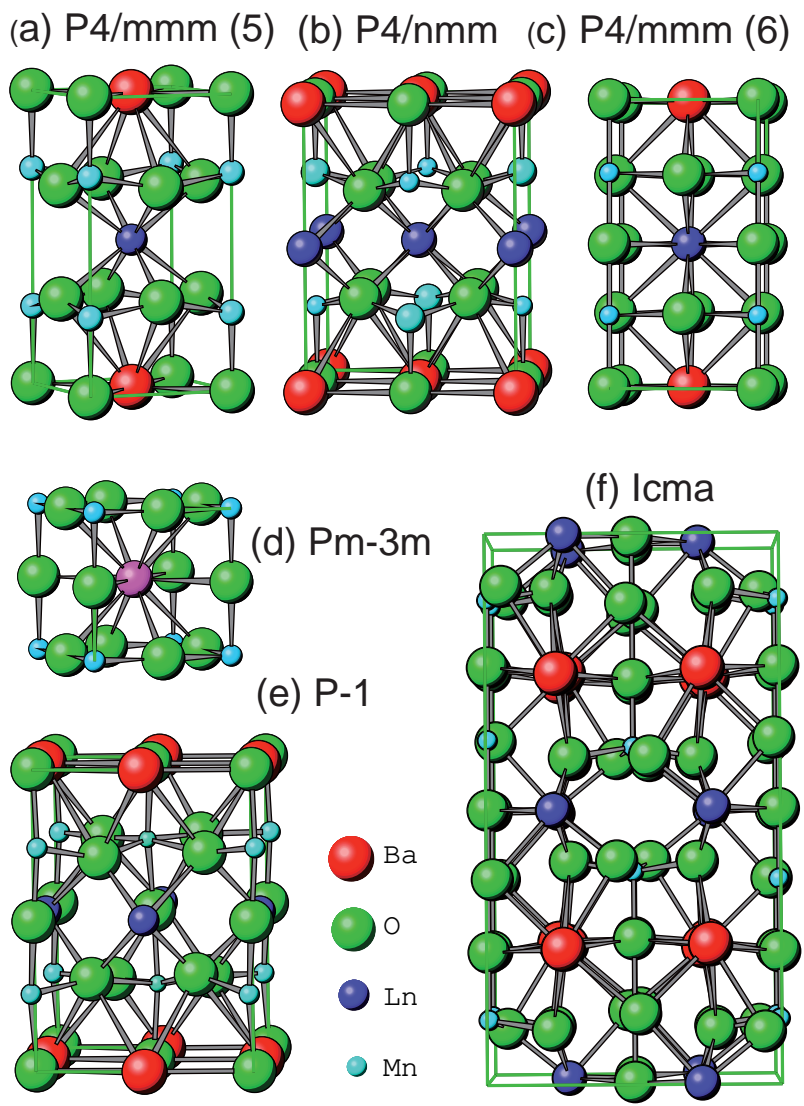}%
\caption{The crystal structures and symmetry groups employed in Rietveld
analysis of x-ray diffraction profiles for layer-ordered LnBaMn$_{2}$O$_{5}%
$(a,b), LnBaMn$_{2}$O$_{6}$(c), disordered Ln$_{0.5}$Ba$_{0.5}$MnO$_{3}$ (d),
YBaMn$_{2}$O$_{6}$(e), and LnBaMn$_{2}$O$_{5.5}$(f). Here Ln = La, Pr, Nd,
(Nd$_{0.9}$Sm$_{0.1}$), (Nd$_{0.1}$Sm$_{0.9}$), Sm and Gd.}%
\label{f2}%
\end{center}
\end{figure}
%EndExpansion

\subsubsection{Oxygen-depleted layered LnBaMn$_{1.96}$Mn$_{0.04}$O$_{5}$.}

We observed that all of our x-ray profiles from "O$_{5}$" samples are
perfectly fitted with the $P4/nmm$ model (Fig.2, b). The quality of fit was
declining when we attempted to fit the patterns in terms of $P4/mmm$ model
(Fig.2, a). This fact is suggestive of persistent charge order which remains
undefeated by the current level of Fe doping. In the pure LaBaMn$_{2}$O$_{5}$,
the COO between Mn$^{2+}$ and Mn$^{3+}$ was initially required from the
occurrence of the supercell extra reflections in electron diffraction
patterns\cite{Millange}. In x-ray profiles, we see only a minor evidence of
COO in the improvement of R-factors for the $P4/nmm$ model. Notable
improvement was observed for every Ln.%

%TCIMACRO{\FRAME{ftbpFU}{4.8402in}{6.877in}{0pt}{\Qcb{ Lattice parameters of
%the reduced perovskite-like cell vs. volume of this cell in 2\% Fe-doped
%manganites. Phases obtained through thermal treatments [Ar-1] (step 1), [O-1]
%(step 2), and [Dis-O] (step 3) are presented in upper panel. In the same
%ranges, phases obtained through thermal treatments [Ar-2] (step 4) and [O-2]
%(step 5) are presented in lower panel. Lattice parameters in two-phase samples
%are plotted versus average cell volume, taking into account the refined
%percentage of each phase. Mixed-rare-earths manganites Sm$_{0.9}$Nd$_{0.1}%
%$BaMn$_{1.96}$Fe$_{0.04}$O$_{y}$ and Sm$_{0.1}$Nd$_{0.9}$BaMn$_{1.96}%
%$Fe$_{0.04}$O$_{y}$ are denoted by "1" and "2", respectively. In YBaMn$_{1.96}
%$Fe$_{0.04}$O$_{6}$ the parameters of the reduced cell are obtained with the
%space group $P2$ and corresponding monoclinoic angle $\beta=$90.296 was taken
%into account in the calculation of the reduced cell volume. }}{\Qlb{f3}%
%}{fi3new.eps}{\special{ language "Scientific Word";  type "GRAPHIC";
%maintain-aspect-ratio TRUE;  display "USEDEF";  valid_file "F";
%width 4.8402in;  height 6.877in;  depth 0pt;  original-width 3.3062in;
%original-height 4.7184in;  cropleft "0";  croptop "1";  cropright "1";
%cropbottom "0";  filename 'Fi3New.EPS';file-properties "XNPEU";}} }%
%BeginExpansion
\begin{figure}
[ptb]
\begin{center}
\includegraphics[
natheight=4.718400in,
natwidth=3.306200in,
height=6.877in,
width=4.8402in
]%
{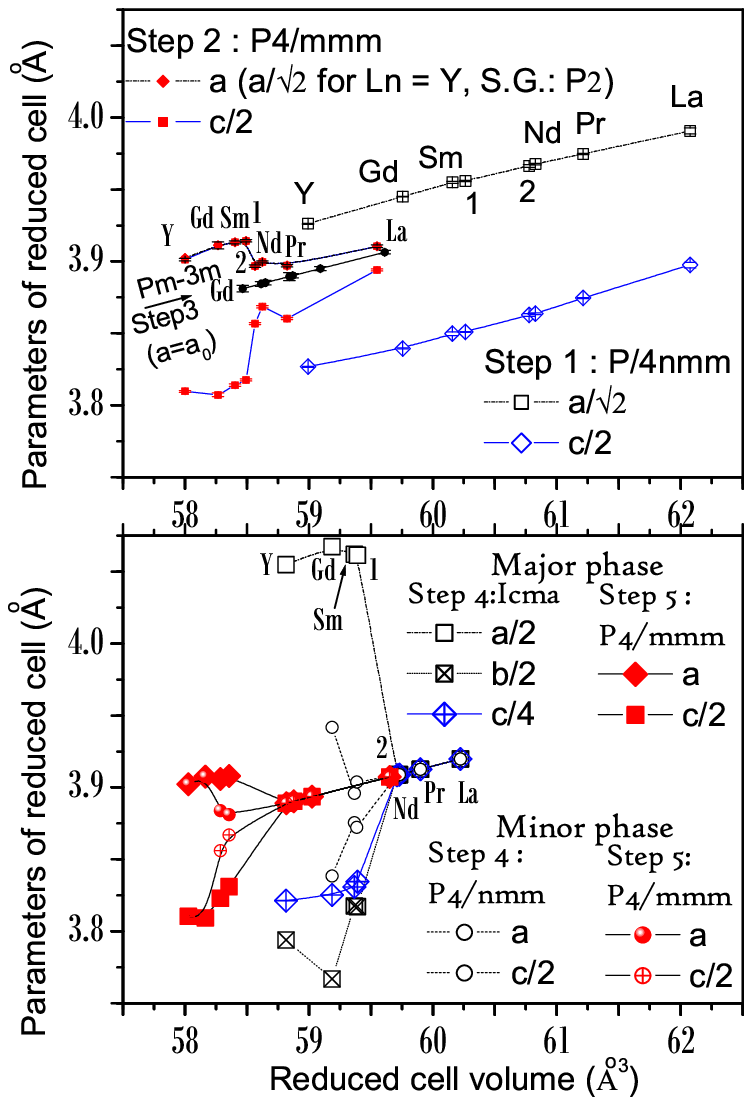}%
\caption{ Lattice parameters of the reduced perovskite-like cell vs. volume of
this cell in 2\% Fe-doped manganites. Phases obtained through thermal
treatments [Ar-1] (step 1), [O-1] (step 2), and [Dis-O] (step 3) are presented
in upper panel. In the same ranges, phases obtained through thermal treatments
[Ar-2] (step 4) and [O-2] (step 5) are presented in lower panel. Lattice
parameters in two-phase samples are plotted versus average cell volume, taking
into account the refined percentage of each phase. Mixed-rare-earths
manganites Sm$_{0.9}$Nd$_{0.1}$BaMn$_{1.96}$Fe$_{0.04}$O$_{y}$ and Sm$_{0.1}%
$Nd$_{0.9}$BaMn$_{1.96}$Fe$_{0.04}$O$_{y}$ are denoted by "1" and "2",
respectively. In YBaMn$_{1.96} $Fe$_{0.04}$O$_{6}$ the parameters of the
reduced cell are obtained with the space group $P2$ and corresponding
monoclinoic angle $\beta=$90.296 was taken into account in the calculation of
the reduced cell volume. }%
\label{f3}%
\end{center}
\end{figure}
%EndExpansion

Comparing the parameters of lattice cell depending on Ln, it is convenient to
plot them in function of the volume $V_{r}$ of the reduced cell, i.e.,
perovskite cell (Fig.2, d). It is shown in Fig.3 that the lattice parameters
in LaBaMn$_{2}$O$_{5}$ vary monotonically as the volume of the reduced cell
increases with increasing the size of Ln. Distortion of the reduced cell can
be calculated in LnBaMn$_{2}$O$_{5}$ as $D_{5}=2(a/\surd2-c/2)/(a/\surd2+c/2)$
. From Y to La the distortion decreases from 2.56\% to 2.36\%, respectively.

\subsubsection{Oxygen-saturated layered LnBaMn$_{1.96}$Mn$_{0.04}$O$_{6}$.}

In contrast to the oxygen-depleted phase, the variation of parameters of the
oxygen-saturated phase vs. $V_{r}$ is non-monotonic. Considering the
distortion of reduced cell $D_{6}=2(a-c/2)/(a+c/2)$ two regions on the $V_{r}
$-axis can be distinguished with $D_{6}\approx2.5\%$ in the region of small
$V_{r}$ and $D_{6}\sim0.7$ to 0.4\% for larger $V_{r}$. (Fig.3). The border
between two regions is the same that divides the Table 1 according to the
products of the treatments [Ar-2] and [O-2].

Contrarily to the case of $y=5$ the quality of x-ray Rietveld refinement was
not improvable with using the enlarged cell $\sqrt{2}a_{\text{p \ }}\times$
\ $\sqrt{2}a_{\text{p \ }}$\ \ $\times$\ \ $2a_{\text{p }}$. The tetragonal
cell $a_{\text{p \ }}\times$ \ $a_{\text{p \ }}$\ \ $\times$\ \ $2a_{\text{p
}}$(Fig.2c) suits best to all the samples in the layer-ordered
oxygen-saturated family, with one exception of YBaMn$_{1.96}$Fe$_{0.04}$%
O$_{6}$. The structure of the latter was refined using the triclinic symmetry
(space group $P\overline{1}$ No.2). Previously, the structure of YBaMn$_{2}%
$O$_{6}$ was described first by the monoclinic symmetry (group $P2$
No.3)\cite{NKIOYU}, however, a recent high resolution diffraction
study\cite{WAR} revealed a triclinic distortion, so that the use of the space
group $P\overline{1}$ is more appropriate, although the resolution in our
profiles was insufficient to confirm the triclinic symmetry. In fact, our
refinements in the Fe-doped samples could not give any preference to one of
the $P\overline{1}$, $P2$ groups.

In members of each of three groups of manganites, as we grouped them by the
size of Ln in the Table 1, a succession of phase transitions was observed via
the magnetization jumps and humps (Fig.4). These features are roughly same as
in undoped manganites, but differ in details. In YBaMn$_{1.96}$Fe$_{0.04}%
$O$_{6}$ and SmBaMn$_{1.96}$Fe$_{0.04}$O$_{6}$, the highest in temperature
jump of magnetzation is associated with the structural transition, which is
triclinic-to-monoclinic for Ln=Y\cite{WAR}, and tetragonal-to orthorhombic for
Ln=Sm\cite{Akah}. Orbital ordering is now believed\cite{UN} to accompany these
structural transition ($T=T_{\text{t}}$), while the complete charge ordering
temperature is attributed\cite{NKU,NKIOYU}, to a separate small hump shifted
from $T_{\text{t}}$ to lower temperature by $\Delta_{\text{t}}$. In
YBaMn$_{1.96}$Fe$_{0.04}$O$_{6}$, for example $\Delta_{\text{t}}\approx40$ K.
In the undoped manganite, the temperature of charge ordering $T_{\text{CO}%
}=T_{\text{t}}-\Delta_{\text{t}} $ is associated with the sharp localization
of charge carriers. In the magnetization of YBaMn$_{1.96}$Fe$_{0.04}$O$_{6}$,
the large jump is observed without any small foregoing hump. Interestingly,
similar disappearance of the hump in magnetization caused by Fe doping was
reported for the orbital ordering transition in BiMnO$_{3}$\cite{Belik}. In a
similar way, the transition is smeared in our Fe-doped samples, and the
magnetization hump is suppressed by the Fe substitution.

Another key feature of the doped systems stems from the fact that the values
of $T_{\text{t}}$ are slightly suppressed compared to those in the undoped
YBaMn$_{2}$O$_{6}$ and SmBaMn$_{2}$O$_{6}$\cite{NKIOYU,NYU}. The suppression
ranges $\Delta T_{\text{t}}($2\%Fe$)$ are of 50K and 40 K for Ln=Y and Sm, respectively.%

%TCIMACRO{\FRAME{ftbpFU}{4.2596in}{6.6691in}{0pt}{\Qcb{Magnetic susceptibility
%$M/H$ measured in the external field $H$ of 1 kOe per mole of formula units in
%LnBaMn$_{1.96}$Fe$_{0.04}$O$_{6}$ for Ln=Y, Sm and (Nd$_{0.9}$Sm$_{0.1}$). The
%zero-fied-cooled magnetization was measured at heating the samples up to
%$T_{\text{max}}$ of 600 K (Ln=Y), 400 K (Ln=Sm) and 370 K (Ln=Nd$_{0.9}%
%$Sm$_{0.1}$) and then at cooling from $T=T_{\text{max}}$. Temperatures of
%phase transitions in pure (undoped) manganites indicated by vertical lines are
%from previous works \cite{NKIOYU,NYU}.}}{\Qlb{f4}}{fi4.eps}%
%{\special{ language "Scientific Word";  type "GRAPHIC";
%maintain-aspect-ratio TRUE;  display "USEDEF";  valid_file "F";
%width 4.2596in;  height 6.6691in;  depth 0pt;  original-width 4.0206in;
%original-height 6.3324in;  cropleft "0";  croptop "1";  cropright "1";
%cropbottom "0";  filename 'Fi4.EPS';file-properties "XNPEU";}} }%
%BeginExpansion
\begin{figure}
[ptb]
\begin{center}
\includegraphics[
natheight=6.332400in,
natwidth=4.020600in,
height=6.6691in,
width=4.2596in
]%
{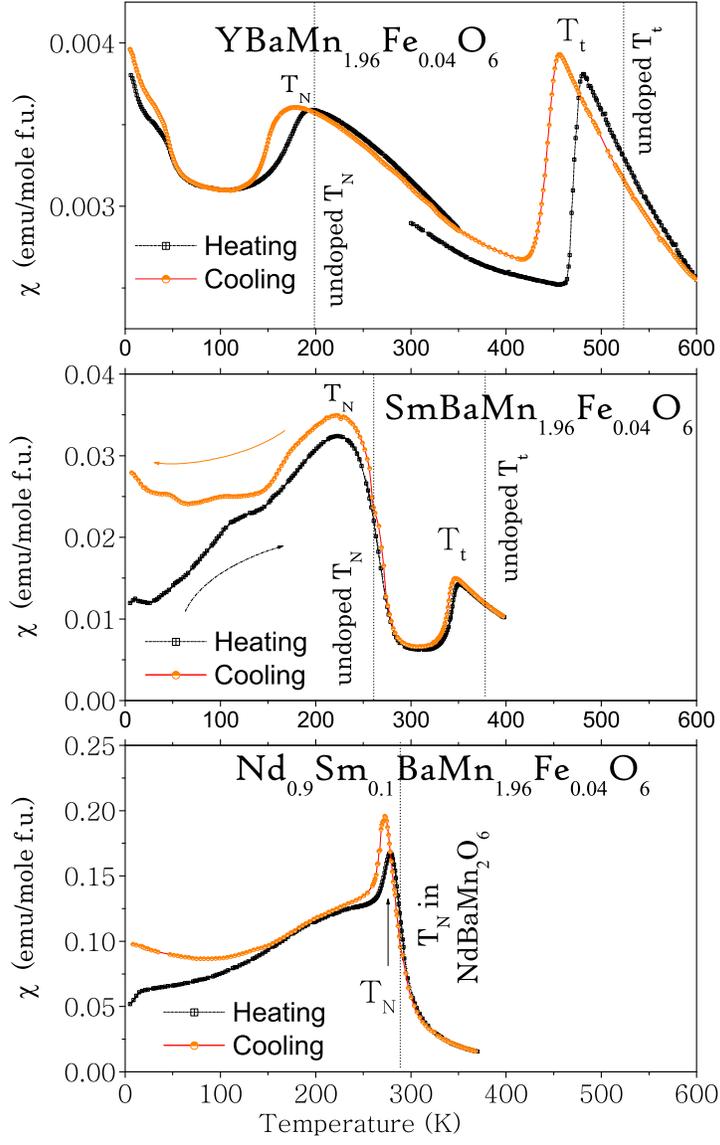}%
\caption{Magnetic susceptibility $M/H$ measured in the external field $H$ of 1
kOe per mole of formula units in LnBaMn$_{1.96}$Fe$_{0.04}$O$_{6}$ for Ln=Y,
Sm and (Nd$_{0.9}$Sm$_{0.1}$). The zero-fied-cooled magnetization was measured
at heating the samples up to $T_{\text{max}}$ of 600 K (Ln=Y), 400 K (Ln=Sm)
and 370 K (Ln=Nd$_{0.9}$Sm$_{0.1}$) and then at cooling from $T=T_{\text{max}%
}$. Temperatures of phase transitions in pure (undoped) manganites indicated
by vertical lines are from previous works \cite{NKIOYU,NYU}.}%
\label{f4}%
\end{center}
\end{figure}
%EndExpansion

Despite of the similarity of the values of $\Delta T_{\text{t}}$, there occurs
a large difference between the cases of Ln=Y and Sm for the shift of the
transition temperature with the reversal of the temperature sweep direction.
Such a shift is associated with an energy barrier for nucleation of a new
phase within the region of overheating or undercooling the preceding phase.
The hysteresis indicates strongly the first-order character of transition that
was observed also in undoped YBaMn$_{2}$O$_{6}$\cite{NKIOYU}. The large
nucleation barrier is observed in YBaMn$_{2}$O$_{6}$ but not in SmBaMn$_{2}%
$O$_{6}$. This is in agreement with a very small structural distortion in
SmBaMn$_{2}$O$_{6}$ at $T_{\text{t}}$ as reported by Akahoshi et al\cite{Akah}.

Temperatures of Neel ($T_{\text{N}}$) are also notably suppressed for both
cases of Ln=Y and Sm, as well as for Nd$_{0.9}$Sm$_{0.1}$BaMn$_{1.96}%
$Fe$_{0.04}$O$_{6}$. The antiferromagnetic transitions humps were
observed\cite{NKYOU,NYU} at 290 K and 250 K in the undoped NdBaMn$_{2}$O$_{6}$
and SmBaMn$_{2}$O$_{6}$, respectively, therefore, the $T_{\text{N}}$\ value of
286 K is expected for the solid solution Nd$_{0.9}$Sm$_{0.1}$BaMn$_{2}$O$_{6}%
$. Remaining suppression $\Delta T_{\text{N}}$ $\simeq10$ K should be
attributed to the effect of Fe substitution. Temperature ranges for
sweep-reversal hysteresis around $T_{\text{N}}$ are not much different between
the three.

\subsubsection{Oxygen-saturated isotropic Ln$_{0.5}$Ba$_{0.5}$Mn$_{0.98}%
$Fe$_{0.02}$O$_{3}$}

According to the Table 1, the disordered phase was obtained in this work from
the ordered one, after the thermal treatment at 1400$^{o}$C in the flowing gas
mixture of 95\% of O$_{2}$ and 5\% of Ar. From the synthesis reported
previously \cite{JAC} our method differed thus by the precursor and gave the
same cubic phases for all Ln's except Y. Only the Y-based complex manganite
has decomposed into binary manganites by this high-temperature oxygenating
treatment. Irrespective of Ln, such a heat treatment was destructive for the
layered arrangement of Ln and Ba formed at the first [Ar-1] step, and resulted
either in isotropic homogeneous distribution of Ln and Ba, or in their
separation for Ln=Y.

A half among 8 compositions studied (in upper part of Table 1) showed a
reduced stability of the disordered phase Ln$_{0.5}$Ba$_{0.5}$Mn$_{0.98}%
$Fe$_{0.02}$O$_{3}$ with respect to a subsequent oxygen-depletion treatment.
The lower stability of the disordered manganites with small Ln correlates with
the poorer perovskite tolerance factor. It is shown in Fig.3 that the lattice
parameters for the Pm$\overline{3}$m phases [Ar-2] obtained after the step 4
are larger than those for the Pm$\overline{3}$m phases [Dis-O] obtained after
the step 3. The differences $\Delta a$ are 0.013, 0.018, and 0.019 \AA \ for
Ln= La, Pr, and Nd, respectively. Increasing $\Delta a$ indicates the
increasing oxygen loss at the step 4. In the [Ar-2] samples, the oxygen index
decreases from La to Nd and approaches the value of 2.75 near the border
separating Pm$\overline{3}$m and Ima2 phases. This is in agreement with the
single-phase [Ar-2] sample of Sm$_{0.9}$Nd$_{0.1}$BaMn$_{1.96}$Fe$_{0.04}%
$MnO$_{5.5}$.

\subsubsection{Intermediate-oxygen LnBaMn$_{1.96}$Mn$_{0.04}$O$_{5.5}$}

The Ima2 phase contains layers BaO and YO$_{0.5}$ and the 1D channels are
formed in the YO$_{0.5}$ layer (Fig. 2,f). In several of the [Ar-2] samples
the second phase (fully oxygen-depleted LnBaMn$_{1.96}$Mn$_{0.04}$O$_{5}$)
appears because the Ima2 structure does not accommodate any oxygen deficiency.
There appears to exist a miscibility gap between the "O$_{5.5}$" and "O$_{5}$"
phases for the Ln in the lower half of the Table 1.

Distortion of the reduced cell can be described by the out-of-plane and
in-plane parameters, having their origins in the layered ordering of Ln and Ba
and the chainlike ordering of oxygen atoms, respectively. The out-of-plane
distortion $D_{5.5}=$ $2(a/4+b/4-c/4)/(a/4+b/4+c/4)$ is just slightly larger
than $D_{5}$ and $D_{6}$. We obtained for Ln=Gd $D_{5.5}=2.77\%$ roughly
similar to $D_{5}$ and $D_{6}$. The in-plane parameter is the orthorhombicity
of the reduced cell $D_{\text{o}}=2(a/2-b/2)/(a/2+b/2).$ For Ln=Gd we obtained
$D_{\text{o}}=$ $6.91\%$, which is much larger than $D_{5.5}$.

\subsubsection{Additional oxygenating thermal treatment [O-2].}

The final thermal treatment [O-2] conducted in oxygen gas flow at rather low
temperature (350$^{o}$C) evidences a peculiar COO-melted phases obtained via
oxygen loading into SmBaMn$_{1.96}$Fe$_{0.04}$O$_{5.5}$ and Sm$_{0.9}%
$Nd$_{0.1}$BaMn$_{1.96}$Fe$_{0.04}$O$_{5.5}$. In the group Ln=La, Pr and Nd,
the lattice parameters before step 4 and after step 5 coincided (Fig.3). Also,
for Ln=Y and Gd one observes (Fig.3) that both lattice parameters after step 5
have regained the values which they had before the step 4. On the other hand,
neither [O-1] nor [Dis-O] x-ray patterns were restituted after the step 5 for
Ln=Sm and Ln=(Sm$_{0.9}$Nd$_{0.1}$). As a matter of fact, the reoxygenation of
[Ar-2] samples for these Ln's resulted in the two phase systems with
approximately equal abundance of each phase. The x-ray profile refined with
two P4/mmm phases gave us the parameters of these phases. For one of the
phases they coincided with the parameters of the [O-1] phase, and for the
other phase the distortion $D_{6}$ was as small as the distortion in the [O-1]
phases for Ln=La, Pr, and Nd. We interpret the latter as the quenched
COO-melted phase and it is confirmed further below via our M\"{o}ssbauer study.

\subsection{M\"{o}ssbauer Study}

In agreement with the structure differences described above among three groups
of Ln-based manganites, our M\"{o}ssbauer study also reveals the
characteristic features in each group. Hereafter, these groups are specified
as La-group, Sm-group and Y-group.

\subsubsection{La-group of Ln's}

The characteristic feature of M\"{o}ssbauer spectra in the [Ar-1] samples
("O$_{5}$"-phase, P4/nmm) is the two-doublet envelope of the spectra. We
observe that the occurrence of two crystallographic sites in this phase for
the largest Ln is sensed by the Fe probe. Both sites accommodate iron in the
state of Fe(III). The ratio of the doublet areas is 7:3 (Fig.5, top) instead
of 5:5 expected for the random distribution of Fe over two sites. This
indicates the iron preference towards one of these sites. The temperature at
which the charge-orbital order sets in is too low to activate the migration of
iron between the Mn(III) and Mn(II) sites. Therefore, we believe that the
dopant species control the in-plane arrangement of the charges and orbitals at
the surrounding ions of Mn. This becomes possible with decreasing the size of
the ordered domain, especially when the COO correlation length becomes
comparable to the average distance between the Fe dopants. The size of the
Mn(II) site is too large for the Fe$^{3+}$ ion; therefore, the Fe$^{3+}%
$species tend to escape to the smaller site of Mn(III). This reslts in the
formation of a static or dynamic configuration of charge ordered domains,
which accommodate the majority of the Fe$^{3+}$ dopants into the Mn(III) site
surrounded by four next-neighboring (NN) Mn$^{2+}$ species through the
in-plane linkage and by one NN Mn$^{3+}$ through the pyramid apex. Less
obvious is the assignment of the secondary doublet can that be attributed
either to Mn(II) site, or to a boundary domain site, surrounded in plane by
both Mn$^{3+}$ and Mn$^{2+}$.%

%TCIMACRO{\FRAME{ftbpFU}{4.4257in}{5.9739in}{0pt}{\Qcb{ Room-temperature
%M\"{o}ssbauer spectra of LaBaMn$_{1.96}$Fe$_{0.04}$O$_{6}$ obtained by the
%sequential [Ar-1], [O-1], [Dis-O], [Ar-2] and [O-2] thermal treatments. The
%conditions of each step are shown in Table 1. }}{\Qlb{f5}}{fi5.eps}%
%{\special{ language "Scientific Word";  type "GRAPHIC";
%maintain-aspect-ratio TRUE;  display "USEDEF";  valid_file "F";
%width 4.4257in;  height 5.9739in;  depth 0pt;  original-width 3.86in;
%original-height 5.2294in;  cropleft "0";  croptop "1";  cropright "1";
%cropbottom "0";  filename 'Fi5.EPS';file-properties "XNPEU";}} }%
%BeginExpansion
\begin{figure}
[ptb]
\begin{center}
\includegraphics[
natheight=5.229400in,
natwidth=3.860000in,
height=5.9739in,
width=4.4257in
]%
{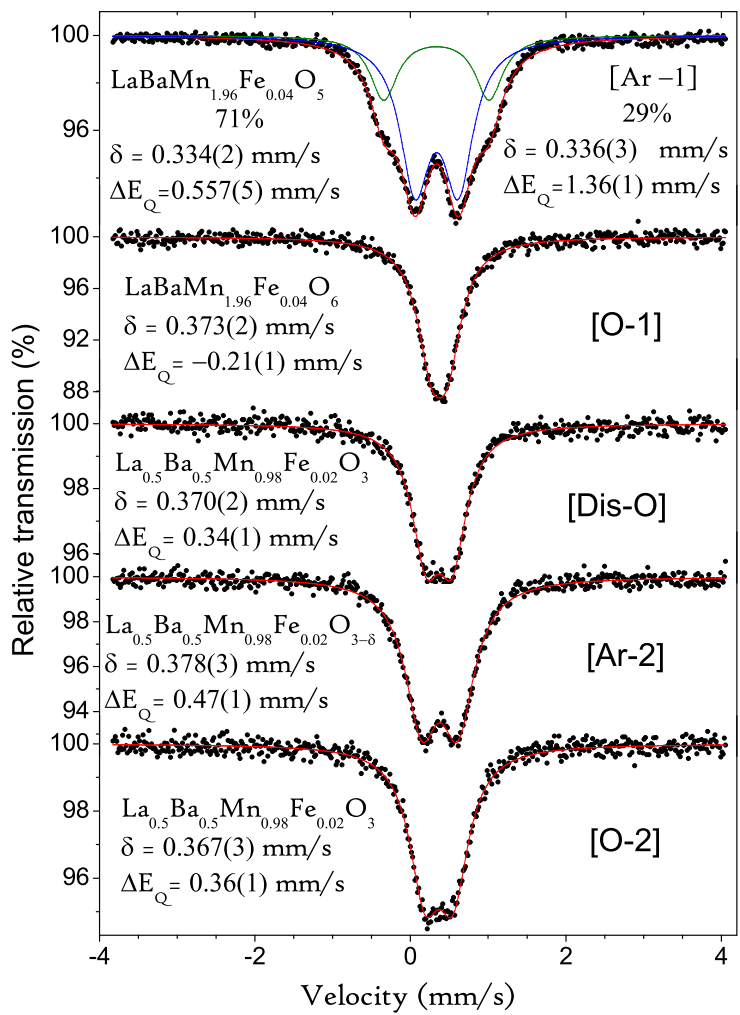}%
\caption{ Room-temperature M\"{o}ssbauer spectra of LaBaMn$_{1.96}$Fe$_{0.04}%
$O$_{6}$ obtained by the sequential [Ar-1], [O-1], [Dis-O], [Ar-2] and [O-2]
thermal treatments. The conditions of each step are shown in Table 1. }%
\label{f5}%
\end{center}
\end{figure}
%EndExpansion

Quadrupole splittings $\Delta E_{Q}$ of these two doublets are in agreement
with our assignment. From the M\"{o}ssbauer studies in cuprates \cite{HI} it
is known that the Fe$^{3+}$ ions doped into the pyramidal site are shifted by
$u$ from the base of the pyramid towards its apex with $u$ much larger for the
Fe dopants than for the host JT ions. When the Fe$^{3+}$ ions were the host
ions instead of the impurities then the displacement parameter $u$ was also
much larger in the FeO$_{5}$ pyramid than in the CuO$_{5}$
pyramid\cite{Er,CaignaertJSSC}. Such an excessive displacement equalizes the
in-plane and the out-of-plane bonds of the Fe$^{3+}$ ions and results in
strong reduction of the splitting $\Delta E_{Q}$ compared to its value
expected from the ionic point charge model for the original pyramidal site of
a JT ion\cite{HI}. We observe for the major doublet in the [Ar-1]-sample of
LnBaMn$_{1.96}$Mn$_{0.04}$O$_{5}$ that both the splitting $\Delta E_{Q}=0.56$
mm/s and the chemical shift $\delta=0.33$ mm/s are quite similar to $\Delta
E_{Q}$ and $\delta$ of Fe doped into Cu(2)-plane of YBa$_{2}$Cu$_{3}$O$_{7-x}$
\cite{HI,Pav,Ry}. This is consistent with our assignment of this doublet to
the site of JT ion Mn$^{3+}$.

Minor doublet has a similar isomer shift, but $\Delta E_{Q}$ as large as 1.36
mm/s. Assignment of this $\Delta E_{Q}$ to the site of Mn(II) is plausible
because this site is too large for Fe$^{3+}$, first of all, in equatorial
dimension\cite{Millange,MSCB}. Embedded in such a site Fe$^{3+}$ cation would
distort its environment or move into an asymmetric position with large $\Delta
E_{Q}$. An alternative assignment of this doublet, which cannot be
disregarded, is a special site, located on the interface between two
charge-ordered domains.

In the oxygen-saturated [O-1] sample, we observe that the value of the
chemical shift $\delta$ is increased up to $0.37$ mm/s showing that the
pyramid completes up to octahedron. With decreasing distortion from $D_{5}$ to
$D_{6}$ the value of $\Delta E_{Q}$ for Ln=La drops down to 0.21 mm/s. The
lack of quadrupole splitting is in line with the absence of COO for this group.

In the disordered La$_{0.5}$Ba$_{0.5}$Mn$_{0.98}$Fe$_{0.02}$O$_{3}$ the value
of $\Delta E_{Q}$ increases again up to 0.34 mm/s. This increase may originate
from the the local strains related to La and Ba randomness. The local strains
are likely to increase after the [Ar-2] treatment when $\Delta E_{Q}$
increases up to 0.47 mm/s. This is caused by some oxygen loss, indicated by
the increased lattice parameter of the cubic Pm$\overline{3}$m phase. Finally,
when the oxygen content "O$_{6}$" was restored after the [O-2] treatment the
splitting $\Delta E_{Q}$ has regained the value characteristic of the [Dis-O] sample.

\subsubsection{Sm-group of Ln's}

In the oxygen-depleted series LnBaMn$_{1.96}$Mn$_{0.04}$O$_{5}$, both
parameters of the lattice vary smoothly as the size of Ln changes (see Fig. 3,
upper panel). However, we observe that the M\"{o}ssbauer spectra change more
drastically as we proceed from La-group to Sm-group. The area of minor doublet
in the uppermost spectrum of Fig.6 is decreased dramatically compared to the
uppermost spectrum of Fig.5. This confirms the spurious nature of the
vanishing minor doublet. If this doublet is assigned to Fe$^{3+}$ in the
Mn(II) site, its disappearance must be associated to the reduction of the COO
correlation length. In this situation, the ordered domains form around the
Fe$^{3+}$ dopants, which serve the anchors to pin the lattice distortions
associated with COO. On the opposite, if the minor doublet originates from the
Fe$^{3+}$ions at the COO grain-boundary site, the vanishing minor doublet
would signify the reduced population of the boundaries and the growing COO
correlation length.%

%TCIMACRO{\FRAME{ftbpFU}{4.4029in}{5.8827in}{0pt}{\Qcb{ Room-temperature
%M\"{o}ssbauer spectra of SmBaMn$_{1.96}$Fe$_{0.04}$O$_{6}$ obtained by the
%sequential [Ar-1], [O-1], [Dis-O], [Ar-2] and [O-2] thermal treatments. The
%conditions of each step are shown in Table 1. }}{\Qlb{f6}}{fi6.eps}%
%{\special{ language "Scientific Word";  type "GRAPHIC";
%maintain-aspect-ratio TRUE;  display "USEDEF";  valid_file "F";
%width 4.4029in;  height 5.8827in;  depth 0pt;  original-width 3.8892in;
%original-height 5.2139in;  cropleft "0";  croptop "1";  cropright "1";
%cropbottom "0";  filename 'Fi6.EPS';file-properties "XNPEU";}} }%
%BeginExpansion
\begin{figure}
[ptb]
\begin{center}
\includegraphics[
natheight=5.213900in,
natwidth=3.889200in,
height=5.8827in,
width=4.4029in
]%
{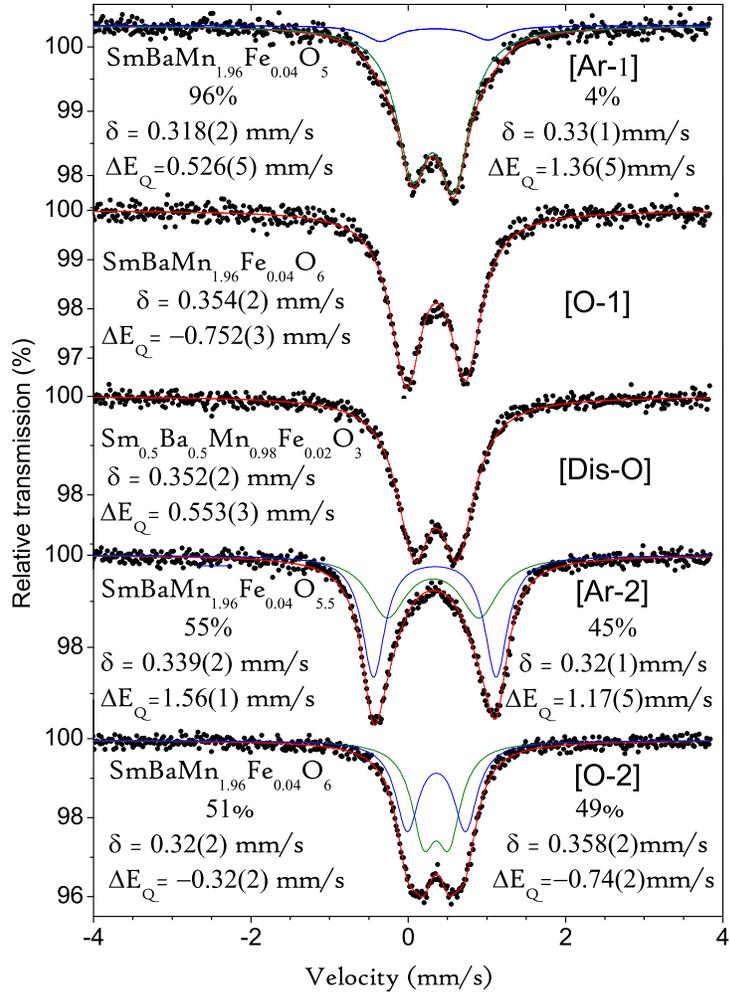}%
\caption{ Room-temperature M\"{o}ssbauer spectra of SmBaMn$_{1.96}$Fe$_{0.04}%
$O$_{6}$ obtained by the sequential [Ar-1], [O-1], [Dis-O], [Ar-2] and [O-2]
thermal treatments. The conditions of each step are shown in Table 1. }%
\label{f6}%
\end{center}
\end{figure}
%EndExpansion

The major doublet appears in an asymmetric form. This asymmetry was
found\cite{subm} to be strongly correlated with the assymetry expected from
the preferred orientation parameters refined using FULLPROF program. This
asymmetry can be removed by applying a special care to randomize the
orientations of platy crystallites at the step of preparation of M\"{o}ssbauer
absorber or by setting the sample at magic angle at the step of measuring the
spectra\cite{subm}. Since this was not always done in this work, the line
intensity ratio of doublets was a free parameter at fitting the spectra.

It is observed more clearly for the Sm-group (Fig. 7) that the relative area
of lines of the main doublet deviates from $1/2$, so that the sign of the
left-to-right line area ratio deviation is changed between the "O$_{5}$" and
"O$_{6}$" series. This is because the main axis of EFG $(z-$axis$)$ is
perpendicular to layers and $V_{zz}>0$ in the pyramid FeO$_{5}$, but
$V_{zz}<0$ in the FeO$_{6}$ octahedron compressed along $z-$axis. Therefore,
in the spectra of both samples [Ar-1] and [O-1](Figs.6,7), the stronger line
is the transition $\pm3/2\rightarrow\pm%
%TCIMACRO{\U{bd}}%
%BeginExpansion
\frac12
%EndExpansion
$, and the weaker line is the transition $\pm%
%TCIMACRO{\U{bd}}%
%BeginExpansion
\frac12
%EndExpansion
\rightarrow\pm%
%TCIMACRO{\U{bd}}%
%BeginExpansion
\frac12
%EndExpansion
$. This is in agreement with the ionic point charge model that prescribes the
EFG sign and orientation along $z$-axis, such that $V_{zz}>0$ for the pyramid
and $V_{zz}<0$ for the compressed octahedron\cite{HI}.

In the "O$_{6}$" state, the charge and orbital order is present for the
Sm-group, but absent for the La-group of Ln's. This is clearly indicated by
the value of $\Delta E_{Q},$which is larger in SmBaMn$_{1.96}$Fe$_{0.04}%
$O$_{6}$ by 3.6 times than in LaBaMn$_{1.96}$Fe$_{0.04}$O$_{6}.$ Large
difference appears also in the values of distortions of the lattice cell,
$D_{6}($Sm$)/D_{6}($La$)=6$. Clearly, the contraction of the cell along the
c-axis originates from the in-plane alignment of the active e$_{g}$ orbitals
of Mn$^{3+}$. The contraction of FeO$_{6}$ octahedron for Ln=Gd is stronger
than that for Ln=Sm and this correspond to $\Delta E_{Q}$ increasing from 0.75
mm/s to 0.85 mm/s(Fig.7).%

%TCIMACRO{\FRAME{ftbpFU}{4.2086in}{5.5588in}{0pt}{\Qcb{ Room-temperature
%M\"{o}ssbauer spectra of GdBaMn$_{1.96}$Fe$_{0.04}$O$_{6}$ obtained by the
%sequential [Ar-1], [O-1], [Dis-O], [Ar-2] and [O-2] thermal treatments. The
%conditions of each step are shown in Table 1. }}{\Qlb{f7}}{fi7.eps}%
%{\special{ language "Scientific Word";  type "GRAPHIC";
%maintain-aspect-ratio TRUE;  display "USEDEF";  valid_file "F";
%width 4.2086in;  height 5.5588in;  depth 0pt;  original-width 3.9339in;
%original-height 5.2139in;  cropleft "0";  croptop "1";  cropright "1";
%cropbottom "0";  filename 'Fi7.EPS';file-properties "XNPEU";}} }%
%BeginExpansion
\begin{figure}
[ptb]
\begin{center}
\includegraphics[
natheight=5.213900in,
natwidth=3.933900in,
height=5.5588in,
width=4.2086in
]%
{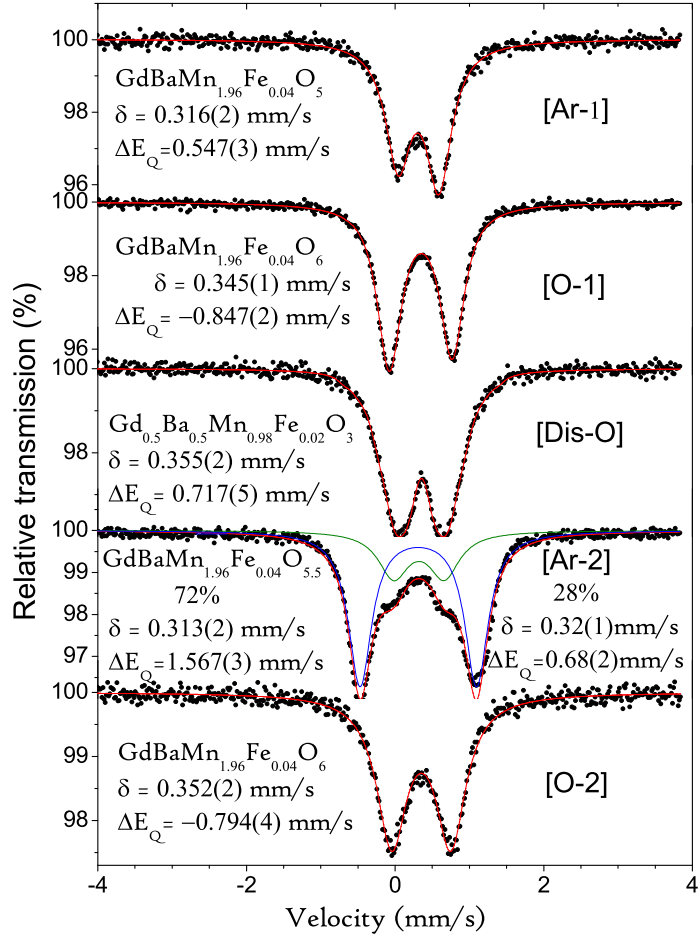}%
\caption{ Room-temperature M\"{o}ssbauer spectra of GdBaMn$_{1.96}$Fe$_{0.04}%
$O$_{6}$ obtained by the sequential [Ar-1], [O-1], [Dis-O], [Ar-2] and [O-2]
thermal treatments. The conditions of each step are shown in Table 1. }%
\label{f7}%
\end{center}
\end{figure}
%EndExpansion

Among the perovskite-like cubic [Dis-O] samples the $\Delta E_{Q}$ value
increases from Ln=La to Sm and further to Gd as 0.37, 0.55, and 0.72 mm/s,
respectively. The value of $\Delta E_{Q}$ is sensing the increase of local
distortion along with the reduction of tolerance factor. In these A-site
disordered manganites, the Lorentzian linewidth parameter is larger than in
the ordered [O-1] samples (cf., for example, 0.52 mm/s and 0.38 mm/s for
Ln=Gd). This evidences the broad distribution of EFG owing to the
inhomogeneity of local strains.

M\"{o}ssbauer spectra of LnBaMn$_{1.96}$Fe$_{0.04}$O$_{5.5}$ were fitted with
two doublets, which could be assigned to the existing sites of Mn(III) with
the octahedral and pyramidal coordinations. Interatomic distances for these
coordinations were reported by Caignaert et al.\cite{Caig} in LaBaMn$_{2}%
$O$_{5.5}$\ and Perca et al.\cite{Perca} in YBaMn$_{2}$O$_{5.5}$. We note that
the geometry of pyramid in YBaMn$_{2}$O$_{5.5}$ is close to a typical one for
the JT Mn$^{3+}$ cation. Two of the in-plane oxygens are at 1.91 \AA , two
others are at 1.935 \AA , and the apical oxygen is at 2.1 \AA . The distortion
in the equatorial plane is much smaller than the apical elongation. The Fe
dopant would be displaced inwards such a pyramid to equalize all five Fe-O
distances. When shifted from the exact crystallographic position of a JT
cation towards the pyramid apex the Fe impurity usually show the moderate
$\Delta E_{Q}$ of the order of 0.6 mm/s\cite{HI}.This is indeed the case of
minor doublet for GdBaMn$_{1.96}$Fe$_{0.04}$O$_{5.5}$ (Fig.7). However, for
the minor doublet in SmBaMn$_{1.96}$Fe$_{0.04}$O$_{5.5}$ the $\Delta E_{Q}$
value of 1.17 mm/s is larger. We attribute this increase of $\Delta E_{Q}$ to
the increased distortion of the pyramid in the equatorial plane. Indeed, in
LaBaMn$_{2}$O$_{5.5}$ Caignaert et al.\cite{Caig} reported the in-plane
interatomic distances of 1.89 and 1.96 \AA .

We assign the major doublet having $\Delta E_{Q}=$1.56 mm/s to the octahedral
coordination, which is strongly elongated along the in-plane $a-$direction
(S.G. Icma). The sign of $\Delta E_{Q}$ is positive, and the principal axis of
EFG coincides with the $a-$axis. Despite of the equal Wyckoff octahedral and
pyramidal position multiplicities, we observe that the octahedral site is more
populated by the Fe dopants than the pyramidal site. Taking into account the
second phase LnBaMn$_{1.96}$Fe$_{0.04}$O$_{5}$ (below 11\% in Table 1) would
only lower the population of the Fe species in the pyramidal site of the main
phase. Unequal populations of these doublets are not counterintuitive because
iron migrates freely between these sites at the temperature of the [Ar-2]
treatment. It must be emphasized that the mechanism of the preference of iron
towards the octahedral site in the [Ar-2] samples is quite different from the
mechanism of the preference of iron towards the Mn(III) site in the COO state
suggested above. Namely, in the case of COO, the iron dopants cannot migrate
at the temperature of the COO formation, and the preference is ensured owing
to the migration of the lattice distortions associated with COO.

Final oxygenating treatment [O-2] resulted in $\Delta E_{Q}=0.8$ mm/s for
GdBaMn$_{1.96}$Fe$_{0.04}$O$_{6}$. This value is not much different from
$\Delta E_{Q}=0.85$ mm/s of [O-1] sample. Concomitantly, one observes a
significant line broadening and a strong decrease of doublet asymmetry. Both
broadening and symmetrization are caused by converting the sample into
nanocrystalline form via oxygenation at the very low temperature of 350$^{o}%
$C. Healing the cracks induced by oxygen intercalation into orthorhombic phase
GdBaMn$_{1.96}$Fe$_{0.04}$O$_{5.5}$ is prevented at such a low temperature.
Orthorhombicity was previously suggested \cite{Taskin} to underlie the strains
and cracks emerging upon the oxygen intercalation into single crystals of
GdBaMn$_{2}$O$_{5.5}$. Nanostructured state of our GdBaMn$_{1.96}$Fe$_{0.04}%
$O$_{6}$ obtained by the [O-2] treatment exhibits the broadening of doublet
due to surface effects in small crystallites. Basal faces of such small
crystallites could be hardly aligned parallel to plane of M\"{o}ssbauer
absorber and the doublet becomes more symmetric.

Two $P4/mmm$ phases observed in x-ray patterns of SmBaMn$_{1.96}$Fe$_{0.04}%
$O$_{6}$ and Sm$_{0.9}$Nd$_{0.1}$BaMn$_{1.96}$Fe$_{0.04}$O$_{6}$ after final
[O-2] treatment manifest themselves also in the M\"{o}ssbauer spectra. Two
doublets were assumed and the linewidths of 0.36 mm/s were obtained for both
doublets having the nearly equal areas and the $\Delta E_{Q}$ values of -0.74
mm/s and -0.32 mm/s. Negative signs of $\Delta E_{Q}$ stand because of
negative $V_{zz}$ in the compressed octahedra. We note that the $\Delta E_{Q}
$ value of the strongly split component coincide with $\Delta E_{Q}$ of the
[O-1] sample. The splitting of weakly split component is close to $\Delta
E_{Q}$ in LaBaMn$_{2}$O$_{6}$ in Fig.5. We note that both $\Delta E_{Q}$ and
the cell distortions $D_{6}$ in PrBaMn$_{1.96}$Fe$_{0.04}$O$_{6}$ and in
NdBaMn$_{1.96}$Fe$_{0.04}$O$_{6}$ are by one and half times larger than
$\Delta E_{Q}$ and $D_{6}$ in LaBaMn$_{1.96}$Fe$_{0.04}$O$_{6}$(see the upper
panel of Fig.3). We attribute the small values of $\Delta E_{Q}$ and $D_{6}$
in the second phase to the absence of of COO. It can be concluded that via
mild oxygenation of SmBaMn$_{1.96}$Fe$_{0.04}$O$_{5.5}$ we succeeded to
stabilize at room temperature (to quench) in SmBaMn$_{1.96}$Fe$_{0.04}$O$_{6}
$ the COO-melted phase. The quenching mechanism is likely related to the
formation of the nanostructured state.

\subsubsection{YBaMn$_{1.96}$Fe$_{0.04}$O$_{6}$}

In YBaMn$_{1.96}$Fe$_{0.04}$O$_{6}$, similarly to SmBaMn$_{1.96}$Fe$_{0.04}%
$O$_{6}$, the M\"{o}ssbauer spectra of the samples [O-1] and [O-2] are
different (Fig.8). The [O-1] sample has showed a single-component spectrum.
Parameters of this doublet at ambient temperature are in line with $\delta$
and $\Delta E_{Q}$ for other Ln's. As the size of Ln decreases in the series
Sm, Gd and Y, the value of $\delta$ remains unaltered, but $\Delta E_{Q}$
varies from --0.75, through --0.85 to --0.97 mm/s, respectively.

On the other hand, the spectrum from [O-2] sample showed a two-doublet nature,
similarly to the [O-2] sample of SmBaMn$_{1.96}$Fe$_{0.04}$O$_{6}$. The
abundance of two components is almost equal and the difference of $\Delta
E_{Q}$ between them is nearly twice (Table 2). The larger $\Delta E_{Q}$
coincides with the quadrupole splitting of the [O-1] sample. Component with a
smaller $\Delta E_{Q}$ is again resembling to that for the quenched
COO-disordered phase.%

%TCIMACRO{\FRAME{ftbpFU}{4.7203in}{3.6991in}{0pt}{\Qcb{M\"{o}ssbauer spectra in
%YBaMn$_{1.96}$Fe$_{0.04}$O$_{6}$ for the [O-1] sample at 295 K, and for the
%[O-2] sample at 295 K and at 11 K.}}{\Qlb{f8}}{fi8.eps}%
%{\special{ language "Scientific Word";  type "GRAPHIC";
%maintain-aspect-ratio TRUE;  display "USEDEF";  valid_file "F";
%width 4.7203in;  height 3.6991in;  depth 0pt;  original-width 5.0339in;
%original-height 3.9325in;  cropleft "0";  croptop "1";  cropright "1";
%cropbottom "0";  filename 'Fi8.eps';file-properties "XNPEU";}} }%
%BeginExpansion
\begin{figure}
[ptb]
\begin{center}
\includegraphics[
height=3.6991in,
width=4.7203in
]%
{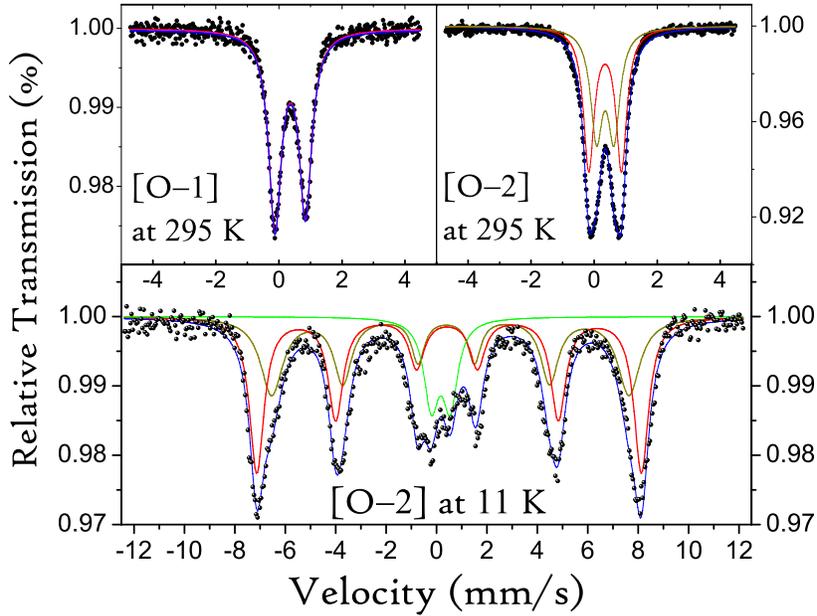}%
\caption{M\"{o}ssbauer spectra in YBaMn$_{1.96}$Fe$_{0.04}$O$_{6}$ for the
[O-1] sample at 295 K, and for the [O-2] sample at 295 K and at 11 K.}%
\label{f8}%
\end{center}
\end{figure}
%EndExpansion

Table 2. Parameters of M\"{o}ssbauer spectra in YBaMn$_{1.96}$Fe$_{0.04}%
$O$_{6}$ at two temperatures: $\delta$ - isomer shift, $\varepsilon-$
quadrupole lineshifts in magnetic sextet*, $\Delta E_{Q}$ - splitting of
doublet in paramagnetic state, H$_{hf}$ - internal magnetic hyperfine field,
$\Theta-$ angle between the principal axis of electric field gradient and
H$_{hf}$, $\Gamma$ and $\Delta\Gamma-$ parameters of linewidth, \% -subspectra area.

\bigskip%

\begin{tabular}
[c]{|l|l|l|l|l|l|l|l|l|l|}\hline
Sam- & T & Subsp- & $\delta$ & $\varepsilon$ & $\Delta E_{Q}$ & H$_{hf}$ &
$\Gamma^{\text{**}}$ & $\Delta\Gamma^{\text{**}}$ & \%\\
ple & (K) & ectrum & mm/s & mm/s & mm/s & kOe & mm/s & mm/s & \\\hline
\lbrack O-1] & 295 & 1 & 0.359(2) & - & -0.967(3) & - & 0.482(4) & - &
100\\\hline
\lbrack O-2] & 295 & 1 & 0.349(2) & - & -1.034(8) & - & 0.410(7) & - & 55\\
&  & 2 & 0.352(2) & - & -0.55(1) & - & 0.45(1) & - & 45\\\hline
\lbrack O-2] & 11 & 1 & 0.463(6) & 0.039(6) &  & 473.1(6) & 0.74(1) &
0.002(3) & 50\\
&  & 2 & 0.47(1) & 0.03(1) &  & 439(2) & 0.54(3) & 0.037(3) & 37\\
&  & 3 & 0.16(1) & - & 0.7(1) & - & 0.7(1) & - & 13\\\hline
\end{tabular}

$^{\text{*}}$The quadrupole lineshift $\varepsilon$ is related to $\Delta
E_{Q}$ via the angle $\Theta$ between $V_{zz}$ and $H_{\text{hf}}$ directions
and the EFG asymmetry parameter $\eta$. See the work \cite{RNUV} for exact
definition of $\varepsilon.$

$^{^{\text{**}}}$Linewidths of six lines of the sextets were fitted with two
parameters $\Gamma$ and $\Delta\Gamma$ assuming the constraints: $\Gamma
_{1,6}=\Gamma+2\gamma_{1}\Delta\Gamma\cdot H/H_{0}$, $\Gamma_{2,5}%
=\Gamma+2\gamma_{2}\Delta\Gamma\cdot H/H_{0}$, $\Gamma_{3,4}=\Gamma
+2\gamma_{3}\Delta\Gamma\cdot H/H_{0}$, where $2\gamma_{1}, $ $2\gamma_{2},$
and $2\gamma_{3}$ are equal to 1, 0.5789555, and 0.157911, respectively, and
$H_{0}=31$ kOe/mm is the constant bringing $H$ into velocity scale
\cite{RJSSC}. The constants $\gamma_{2},$ and $\gamma_{3}$ are given by
$(g_{1/2}-g_{3/2})/(g_{1/2}-3g_{3/2})$ and ($g_{1/2}+g_{3/2})/(g_{1/2}%
-3g_{3/2})$, where $g_{1/2}=0.181208$, $g_{3/2}=-0.103542$ are the g-factors
of ground and excited states of $^{57}$Fe \cite{RNUV}.

\bigskip

M\"{o}ssbauer spectrum taken at $T=11$K in the [O-2] sample of YBaMn$_{1.96}%
$Fe$_{0.04}$O$_{6}$ showed that a part (13\%) of $^{57}$Fe species remained in
paramagnetic state. Magnetically split components show the abundance ratio
similar to the ratio of areas of two doublets at 295 K. From the
relationship\cite{RNUV,RJSSC} between $\Delta E_{Q}$ and $\varepsilon$ one can
specify the angle between $V_{zz}$ and $H_{\text{hf}}$ directions. Assuming
symmetric EFG we obtain the $\Theta$ values of 58(1) and 59(2) degrees for the
components 1 and 2 in Table 2, respectively. However, when a significant
asymmetry parameter $\eta\sim0.8$ is allowed the $\Delta E_{Q}$ vs
$\varepsilon$ relationship results in $\Theta=90^{\text{o}}$. Interatomic
distances shown by Nakajima et al in Fig. 7 of their work\cite{NKIOYU} suggest
indeed that the EFG could be quite asymmetric. Taking the axis of major
compression of the octahedra along c-axis as the principal axis of EFG,
$V_{zz}<0$, we conclude that the $H_{\text{hf}}$ axis lies close to one of
in-plane directions.

\section{Concluding Remarks}

Room-temperature M\"{o}ssbauer spectra in layered manganites evidence the high
sensitivity of the electric field gradient to fine tuning of structure
parameters. Although the substitution of iron for manganese suppresses the
transition temperatures in each of the families of manganites, the consecution
of the phase transformations is preserved. Shortening the COO correlation
length is conjectured to explain the single-component spectra for the
multisite structures. Indeed, Shannon ionic radii for Mn$^{3+}$ and Fe$^{3+}$
coincide exactly (0.645 \AA ), while the radius of Fe$^{3+}$ is too small or
too big for the sites of Mn(II) and Mn(IV), respectively. The fact that we
frequently see the dopants in the site of Mn(III) but rarely anywhere else
suggests the limited involvement of iron into the manganite electronic system.
Iron stands alone not only because of stable valence but also owing to the
tendency of Fe$^{3+}$ ions to accommodate a less distorted environment matched
to the isotropic configuration $d^{5}$. On the other hand, in the octahedral
environment, the Fe$^{3+}$ dopants probe faithfully the distortion of the
lattice. M\"{o}ssbauer spectroscopy observes the fourfold increase of the
electric field gradient in the charge and orbitally ordered phases compared to
the unordered ones. Finally, we proposed that the electronically unordered
system can be quenched to room temperature in the nanocrystalline phase,
although such a quenching was observed mainly in the two-phase samples, where
the quenched melted phase coexisted with the conventional ordered phase.

\section{Acknowledgements}

This work was supported by Asahi Glass Foundation and RFBR-JSPS joint project
(Grant 07-02-91201).

\section{Figure Captions}

Fig. 1. X-ray diffraction patterns of Sm$_{0.9}$Nd$_{0.1}$BaMn$_{1.96}%
$Fe$_{0.04}$O$_{y}$, obtained by thermal treatments [Ar-1], [O-1], [Dis-O],
[Ar-2] and [O-2] explicated in Table 1. Dots represent the observed profiles;
solid lines represent calculated profiles and difference.

Fig.2. The crystal structures and symmetry groups employed in Rietveld
analysis of x-ray diffraction profiles for layer-ordered LnBaMn$_{2}$O$_{5}%
$(a,b), LnBaMn$_{2}$O$_{6}$(c), disordered Ln$_{0.5}$Ba$_{0.5}$MnO$_{3}$ (d),
YBaMn$_{2}$O$_{6}$(e), and LnBaMn$_{2}$O$_{5.5}$(f). Here Ln = La, Pr, Nd,
(Nd$_{0.9}$Sm$_{0.1}$), (Nd$_{0.1}$Sm$_{0.9}$), Sm and Gd.

Fig.3. Lattice parameters of the reduced perovskite-like cell vs. volume of
this cell in 2\% Fe-doped manganites. Phases obtained through thermal
treatments [Ar-1] (step 1), [O-1] (step 2), and [Dis-O] (step 3) are presented
in upper panel. In the same ranges, phases obtained through thermal treatments
[Ar-2] (step 4) and [O-2] (step 5) are presented in lower panel. Lattice
parameters in two-phase samples are plotted versus average cell volume, taking
into account the refined percentage of each phase. Mixed-rare-earths
manganites Sm$_{0.9}$Nd$_{0.1}$BaMn$_{1.96}$Fe$_{0.04}$O$_{y}$ and Sm$_{0.1}%
$Nd$_{0.9}$BaMn$_{1.96}$Fe$_{0.04}$O$_{y}$ are denoted by "1" and "2",
respectively. In YBaMn$_{1.96}$Fe$_{0.04}$O$_{6}$, the plotted parameters of
the reduced cell are obtained using the space group $P2$ and corresponding
monoclinoic angle $\beta=$90.296 was taken into account in the calculation of
the reduced cell volume.

Fig.4. Magnetic susceptibility $M/H$ measured in the external field $H$ of 1
kOe per mole of formula units in LnBaMn$_{1.96}$Fe$_{0.04}$O$_{6}$ for Ln=Y,
Sm and (Nd$_{0.9}$Sm$_{0.1}$). The zero-fied-cooled magnetization was measured
at heating the samples up to $T_{\text{max}}$ of 600 K (Ln=Y), 400 K (Ln=Sm)
and 370 K (Ln=Nd$_{0.9}$Sm$_{0.1}$) and then at cooling from $T=T_{\text{max}%
}$. Temperatures of phase transitions in pure (undoped) manganites indicated
by vertical lines are from previous works \cite{NKIOYU,NYU}.

Fig.5. Room-temperature M\"{o}ssbauer spectra of LaBaMn$_{1.96}$Fe$_{0.04}%
$O$_{6}$ obtained by the sequential [Ar-1], [O-1], [Dis-O], [Ar-2] and [O-2]
thermal treatments. The conditions of each step are shown in Table 1.

Fig.6. Room-temperature M\"{o}ssbauer spectra of SmBaMn$_{1.96}$Fe$_{0.04}%
$O$_{6}$ obtained by the sequential [Ar-1], [O-1], [Dis-O], [Ar-2] and [O-2]
thermal treatments. The conditions of each step are shown in Table 1.

Fig.7. Room-temperature M\"{o}ssbauer spectra of GdBaMn$_{1.96}$Fe$_{0.04}%
$O$_{6}$ obtained by the sequential [Ar-1], [O-1], [Dis-O], [Ar-2] and [O-2]
thermal treatments. The conditions of each step are shown in Table 1.

Fig.8. M\"{o}ssbauer spectra in YBaMn$_{1.96}$Fe$_{0.04}$O$_{6}$ for the [O-1]
sample at 295 K, and for the [O-2] sample at 295 K and at 11 K.

\label{}
\end{document}